\documentclass[prd,twocolumn,graphicx,print]{revtex4-1}
\usepackage{graphicx}
\usepackage{dcolumn}
\usepackage{bm}
\usepackage{epsfig}
\usepackage{lipsum}
\usepackage{natbib}
\usepackage[fleqn]{amsmath}
\usepackage{slashed}
\usepackage{amssymb}
\usepackage{ mathrsfs}
\usepackage{times}
\usepackage{psfrag}
\usepackage[english]{babel}
\usepackage[normal]{subfigure}
\usepackage{multirow}
\usepackage{color}
\usepackage{physics}
\usepackage{changepage}
\usepackage{lipsum}
\begin{document}

\title{A Proposal for a Covariant Entropy Relation}

\author{Dor Gabay}
\email[]{dorgabay@mail.tau.ac.il}
\affiliation{Department of Physical Electronics, Tel-Aviv University, Tel-Aviv 69978, Israel.}

\begin{abstract}
A density-dependent conformal killing vector (CKV) field is attained from a conformally transformed action composed of a unique constraint and a Klein-Gordon field. 
The CKV is re-expressed into an information identity and studied in its integro-differential form for both null and time-like geodesics. 
It is conjectured that the identity corresponds to a generalized second law of thermodynamics which holographically relates the covariant entropy contained within a volumetric $n$- and $(n-1)$-form, starting from an $(n-2)$-spatial area. 
The time-like geodesics inherit an effective `geometric spin' while the null geodesics are suggested to obey the generalized covariant entropy bound so long as they conform to Einstein's equation of state. 
To then comply with the equation of state, a metriplectic system is introduced, whereby a newly defined energy functional is derived for the entropy.
Such an entropy functional mediates the Casimir invariants of the Hamiltonian and therefore preserves the symplectic form of quantum mechanics. 
For null geodesics, the Poisson bracket of the entropy functional with the Hamiltonian is shown to elegantly result in Einstein's energy-mass relation. 
\end{abstract}

\date{\today}


\maketitle 

\section{Introduction}
Even in the midst of total chaos, nature constantly reveals its tendency to find a balance. It is our ability to identify these mechanisms of conservation that govern whether we are able to escalate our understanding of nature. In the past few decades, entropy has been gaining increased attention, in particular pertaining to the area law, first recognized by Bekenstein in his acclaimed work on black holes~\cite{bekenstein1972arealaw}. He demonstrated that the black hole entropy is equal to the ratio of the black hole area to the Planck length squared times a proportionality factor of four. This `generalized' second law motivated Hawking~\cite{hawking1976thermal} to conjecture that the nonzero entropy must result in a loss of information in the form of thermal radiation at some finite temperature. A question then naturally followed: What fundamental principle does such behavior of information impose on the structure of our spacetime and on the governing principles of quantum mechanics?

Since then, theorists attempted to extend the generalized second law to entropy bounds of covariant form. One such conjecture is the Covariant Entropy Bound (CEB) of Bousso~\cite{buosso1999CEB}, whereby the entropy flux of null geodesics, starting from an arbitrary spatial area $A$ and traversing along the volumetric $(n-1)$-form of the corresponding null hypersurface, were hypothesized to be bound by Bekenstein's black hole entropy. Utilizing Raychaduri's equation~\cite{kar2007raychauduri}, this conjecture was rigorously proved and thereafter nicknamed the generalized CEB~\cite{flanagan2000CEB,bousso2003CEB}. It should be recognized that the CEB inequality only provided a bound for the entropy flux of null geodesics. Generalizing this statement to a relation that analogously dictates the covariant entropy of null and time-like geodesics is of prior importance.

Quantum field theory (QFT) is not governed by a second law of thermodynamics. Although the first law is enforced, via Noether conservation~\cite{banados2016noether}, there is no extension of QFT, that I know of, which incorporates additional degrees of freedom to mediate information in correspondence to some fundamental identity. Metriplectic systems~\cite{kaufman1984MP,morrison1986MP,morrison2009MP,materassi2016MP} serve as interesting candidates to do exactly that. They attribute the additional degrees of freedom to the Casimir invariants of the system, in turn conserving the system's symplectic behavior. In that respect, they allow for the incorporation of both a first and second laws of thermodynamics. 

Some fascinating work, along the line of AdS-CFT duality, has been done in mapping the first law of thermodynamics to linear theories of gravity. Faulkner et al.~\cite{faulkner2014entgl} has shown, using the the Ryu-Takenagi formula in AdS spacetime, that entanglement entropy articulates a quantum generalization of the first law of thermodynamics. More specifically, that the dual space of enclosed areas in conformal field theories, dwelling on the surface of AdS spacetimes, are no more then an exact representation of linear theories of gravity in the corresponding bulk structure. 
Since then, many works~\cite{lashkari2014entgl,swingle2014entgl,jacobson2016entgl} have dealt with the injective mapping of entanglement entropy of conformal field theories, characterized by the Von-Neumann prescription, to linear gravity in AdS-spacetime. 
Jacobson~\cite{jacobson2016entgl} identified that for a spatial ball of fixed volume, one attains a correspondence between the variation in area characterized by Einstein's equation and the `maximum' entanglement entropy of conformal field theories. This can only be true by accounting for the fact that the modular Hamiltonian generates the flow of a conformal killing vector field, which, to the best of my knowledge, is only present in conformal field theories. This equivalence was nicknamed the `maximum vacuum entanglement entropy.' 

I conjecture that a more general principle must then be incorporated into QFT in the form of a holographic principle, one that naturally accounts for a balance between the covariant entropies contained within the volumetric $(n-1)$- and $n$-forms starting from some enclosed $(n-2)$-spatial area $A$, and that such a conservation mechanism could be enforced by an identity taking the form of a density-dependent CKV. 

A proper extension of quantum mechanics would then have to dwell beyond locality, inevitably resulting in a finite UV-spectrum. 
Within a conformally transformed action, a constraint can be used to enforce the conformal factor to conform with the effective mass of the Hamilton-Jacobi equation~\cite{our_cosmo_paper}. 
The constraint would allow the non-conserved Noether current to be mediated by its Lagrange multiplier $\lambda$ so that, in some limit, QFT's intrinsic locality can be retrieved by this additional degree of freedom. 
Furthermore, the constraint can be used to introduce a density-dependent CKV which naturally establishes a hologrophic correspondence for information. 
Within the framework of metriplectic systems, such a theory can be shown to naturally imply first and second laws of thermodynamics for both null and time-like geodesics. 

I begin by providing a brief introduction to the theory under consideration~\cite{our_cosmo_paper,gabay2020eom}, followed by a derivation of the aforementioned density-dependent CKV. Thereafter, the CKV is re-expressed into a holographic form and conjectures for null and time-like geodesics are presented. For null hypersurfaces, it is further suggested that for the holographic identity to conform with the well established CEB, one must obey Einstein's equation of state. For the proposed theory to then conform with the equation of state, a metriplectic system is explored, resulting in an additional energy functional for the entropy. The Poisson bracket of the Hamiltonian with the newly defined entropy functional is derived. Finally, in the limit of null geodesics, it is demonstrated that such a Poisson bracket elegantly results in Einstein's energy-mass relation.

\section{Theory}
Recently, a certain class of constrained conformally transformed actions~\cite{our_cosmo_paper}, expressed using the Bohmian interpretation of quantum mechanics, were suggested to bypass Weinberg's 1989 no-go theorem~\cite{weinberg1989nogo}. The linear-order conformal factor $\Omega_{lin}^{2}$, used to conformally transform the action, was chosen to comply with the fluctuations of the effective mass of the relativistic Hamilton-Jacobi equation for a Klein-Gordon field
\begin{eqnarray}
\nabla_{\mu}S\nabla^{\mu}S=\Omega_{lin}^{2}m^{2}=(1+Q)m^{2}. \label{HJeqn_lin}
\end{eqnarray}
Here, $Q=\frac{\Box\sqrt{\rho}}{\sqrt{\rho}}$ is the relativistic quantum potential~\cite{Bohm1975,BohmI,BohmII}. $S$ and $\rho$ are the phase and density, respectively, associated to the scalar field $\varphi=\sqrt{\rho}\textup{e}^{iS/\hbar}$. To remove any ghost (energy) states present to linear-order, the Hamilton-Jacobi equation is extended to its exponential form 
\begin{equation}
\begin{aligned}
\nabla_{\mu}S\nabla^{\mu}S&=\Omega^{2}m^{2} \\
&=\textup{e}^{Q}m^{2}=(1+Q+\frac{1}{2}Q^{2}+...)m^{2}. \label{HJeqn}
\end{aligned}
\end{equation}
The scalar field representation of the Lagrangian (with metric signature $(1,-1,-1,-1)$), along with the prescribed constraint, is given by
\begin{widetext}
\begin{eqnarray}
A[g_{\mu\nu},\Omega,\varphi,\lambda]=\int\mathrm{d}^{4}x\sqrt{-g}\Biggl\{\mathcal{L}_{EH}(g_{\mu\nu},\Omega)+\frac{\Omega^{2}}{2}\Bigl[\nabla_{\mu}\varphi^{*}\nabla^{\mu}\varphi-\Omega^{2}\frac{m^{2}}{\hbar^{2}}\varphi^{*}\varphi\Bigr]+\lambda\Bigl[\mathrm{ln}(\Omega^{2})-\mathcal{G}(\varphi,\varphi^{*},\nabla_{\mu}\varphi,\nabla_{\mu}\varphi^{*},...)\Bigr]\Biggr\}. \nonumber \\ \label{vacuaA}
\end{eqnarray}
\end{widetext}
Here, $\mathcal{L}_{EH}$ is the conformally transformed Einstein-Hilbert Lagrangian and the functional $\mathcal{G}$ is the scalar field representation of the relativistic quantum potential $Q$. The derivation for the equations of motion which follow can be found in previous works~\cite{our_cosmo_paper,gabay2020eom}. 

In ensuring the scalar and tensor degrees of freedom align, the trace of the Lagrangian, varied with respect to the metric tensor $g_{\mu\nu}$, is set equal to the Lagrangian varied with respect to the conformal factor 
\begin{eqnarray}
\frac{\delta\mathcal{L}}{\delta\Omega}=\textup{Tr}\Biggl(\frac{\delta\mathcal{L}}{\delta g^{\mu\nu}}\Biggr). \label{Lvar}
\end{eqnarray}
The resulting equation of motion for the dynamical field $\lambda$, in its compact representation, takes the form
\begin{eqnarray}
\frac{m^2}{\hbar^2}(1-Q)\lambda=\nabla_{\mu}\Bigl(\lambda\frac{\nabla^{\mu}\sqrt{\rho}}{\sqrt{\rho}}\Bigr). \label{Slambda}
\end{eqnarray}
The equation of motion governing the Lagrange multiplier $\lambda$, hereon referred to as the mediating field, resembles a diffusion equation~\cite{gabay2020eom}. This equation is intrinsically coupled to the scalar field's corresponding wave equation, mediating its newly added source and dissipation terms. Unlike the linear conformal factor $\Omega^{2}_{lin}$ of Eq.~(\ref{HJeqn_lin}), the exponential conformal factor of Eq.~(\ref{HJeqn}) results in bound, nonsingular solutions of $\lambda$~\cite{our_cosmo_paper}.

In the limit of $\lambda\to\rho$, hereon referred to as the field alignment scenario, the prototypical Klein-Gordon equation is retrieved, along with its Poincar$\acute{\textup{e}}$ invariance. Furthermore, the mass is alleviated from the scalar field equation, resulting in a local conformal invariance. This only occurs when the conservation conditions of the scalar and mediating fields are satisfied~\cite{gabay2020eom}. 

A tensor representation of the equation of motion for $\lambda$ can similarly be attained by instead metricating the left-hand side of Eq.~(\ref{Lvar}) 
\begin{eqnarray}
\nabla_{(\mu}\Bigl(\frac{\lambda}{\sqrt{\rho}}\Bigr)\nabla_{\nu)}\sqrt{\rho}-g_{\mu\nu}\nabla_{\sigma}\Bigl(\lambda\frac{\nabla^{\sigma}\sqrt{\rho}}{\sqrt{\rho}}\Bigr)=\frac{m^2}{\hbar^2}\lambda g_{\mu\nu}. \label{Tlambda}
\end{eqnarray}
Here, the parenthesis placed within the index notation are symbolic of the symmetrized permutation of the indices. 
We further attempt to merge these two expressions in their tensor form. In first multiplying Eq.~(\ref{Slambda}) by $g_{\mu\nu}$, followed by summing over the permutation of the resulting tensor equation, one arrives at the following symmetric tensor relation
\begin{eqnarray}
\nabla_{(\mu}\Bigl(\frac{\lambda}{\sqrt{\rho}}\Bigr)\nabla_{\nu)}\sqrt{\rho}=\frac{2m^2}{\hbar^2}\lambda g_{\mu\nu}-2\frac{\lambda}{\sqrt{\rho}}\nabla_{(\mu}\nabla_{\nu)}\sqrt{\rho}. \nonumber \\
\end{eqnarray}
The above relation can then be substituted into Eq.~(\ref{Tlambda}) and, after some algebraic manipulation, the following $\lambda$-independent identity can be retrieved
\begin{eqnarray}
\nabla_{(\mu}\nabla_{\nu)}\sqrt{\rho}=\frac{1}{2} g_{\mu\nu}\nabla_{\sigma}\nabla^{\sigma}\sqrt{\rho}. \label{CKVlambda_v0}
\end{eqnarray}
Although not obvious at first glance, the resulting expression is simply a CKV. The Killing vector $K_{\mu}=\nabla_{\mu}\sqrt{\rho}$ is representative of the spatio-temporal variations of the density $\sqrt{\rho}$. The physical meaning of this is not entirely intuitive at first, given that the majority of Killing vector fields in field theory articulate diffeomorphisms associated to phase space (i.e. Penrose operators~\cite{penrose1967poper,penrose1984poper}). A brief introduction to CKVs, along with their relation to Killing Tensor (KT) fields is provided in Appendix~\ref{sec:app1}. 

The CKV of Eq.~\ref{CKVlambda_v0} holds so long as the scalar and mediating fields are not aligned. In the limit of the field alignment scenario ($\lambda\to\rho$), the mediating field ensures a Noether current conservation and the relativistic quantum potential begins to nullify $Q\approx0$~\cite{gabay2020eom}. In such a limit, the massless Klein-Gordon field results in a nearly uniform particle density. The equation of motion for the mediating field would then drastically simplify
\begin{eqnarray}
\frac{m^2}{\hbar^2}\lambda=\frac{1}{2}\Box\lambda. \label{Slambda_s}
\end{eqnarray}
In similarly merging this expression with that of its tensor form in Eq.~(\ref{Tlambda}) (within the field alignment scenario), a Killing vector (KV) field can then be retrieved
\begin{eqnarray}
\nabla_{(\mu}\nabla_{\nu)}\sqrt{\rho}=0. \label{KVlambda}
\end{eqnarray}
It can therefore be concluded that the CKV obeys a KV for null geodesics. In deciphering the CKV of Eq.~(\ref{CKVlambda_v0}), the density can be interpreted as characterizing the intrinsic behavior of information in spacetime. One possible approach would be to define its associated entropy relation and, in what follows, study its null and time-like geodesic properties. My hope is that such an expression would result in the proper characterization of the generalized second law of thermodynamics for both quantum and classical phenomena.

\section{An Entropy Relation}
Whether one adopts the Shannon, Von Neumann, or Gibbs entropies, they all tend to characterize one uncontroversial aspect of nature: a measure of the disorder of information. Each of these entropies contains intricately different properties and cannot be deemed to be the same. 
Unlike other measures of entropy, the Von Neumann entropy $\mathfrak{S}=-\textup{Tr}[\rho\ln\rho]$ respects the property of sub-additivity, so as to conform with the non-commutative nature of quantum mechanics, and will henceforth be adopted. Given we are currently interested in the dynamics of a single field, it suffices to remove the trace for a scalar density $\mathfrak{S}=-\rho\ln\rho$.

To better shine light at the CKV of Eq.~(\ref{CKVlambda_v0}), we re-express it in a $\rho$-dependent form. Using the relation $\nabla_{\mu}\rho/\rho=2\nabla_{\mu}\sqrt{\rho}/\sqrt{\rho}$ and accounting for the fact that 
\begin{eqnarray}
\frac{\nabla_{\sigma}\nabla^{\sigma}\sqrt{\rho}}{\sqrt{\rho}}=\frac{1}{2}\nabla_{\sigma}\Bigl(\frac{\nabla^{\sigma}\rho}{\rho}\Bigr)+\frac{1}{4}\frac{\nabla_{\sigma}\rho}{\rho}\frac{\nabla^{\sigma}\rho}{\rho}, \label{CKVlambda_v1}
\end{eqnarray}
one can obtain a $\rho$-dependent, rather than $\sqrt{\rho}$-dependent, identity
\begin{eqnarray}
\Biggl(\nabla_{(\mu}&&\Bigl(\frac{\nabla_{\nu)}\rho}{\rho}\Bigr)+\frac{1}{2}\frac{\nabla_{(\mu}\rho}{\rho}\frac{\nabla_{\nu)}\rho}{\rho}\Biggr)= \nonumber \\
&&\frac{2}{n}g_{\mu\nu}\Biggl(\nabla_{\sigma}\Bigl(\frac{\nabla^{\sigma}\rho}{\rho}\Bigr)+\frac{1}{2}\frac{\nabla_{\sigma}\rho}{\rho}\frac{\nabla^{\sigma}\rho}{\rho}\Biggr). \label{CKVlambda_v2}
\end{eqnarray}
For practical purposes, we can begin by expressing the change in information associated the particle density $\nabla_{\mu}S=-\nabla_{\mu}(\ln\rho)=-\nabla_{\mu}\rho/\rho$ so as to arrive to the following relation 
\begin{eqnarray}
\nabla_{(\mu}\nabla_{\nu)}&&{S}-\frac{1}{2}\nabla_{(\mu}{S}\nabla_{\nu)}{S} \nonumber \\
&&=\frac{2}{n}g_{\mu\nu}\Bigl[\nabla_{\sigma}\nabla^{\sigma}S-\frac{1}{2}\nabla_{\sigma}S\nabla^{\sigma}S\Bigr]. \label{Sinfo}
\end{eqnarray}
Although this expression clearly defines an identity for the measure of information, it is pertinent to more rigorously define the CKV of Eq.~(\ref{CKVlambda_v0}) in its Von Neumann form. This can be done by utilizing the following relationship
\begin{eqnarray}
\frac{\nabla_{\mu}\rho}{\rho}=\frac{\nabla_{\mu}(\rho\ln\rho)}{\rho+\rho\ln\rho}. \label{VN_v0}
\end{eqnarray}
This relation can be quickly verified by applying a chain rule on the left-hand side. In substituting Eq.~(\ref{VN_v0}) into the bracketed expression on the left-hand side of Eq.~(\ref{CKVlambda_v2}), one arrives at the following 
\begin{eqnarray}
\Biggl(\nabla_{\sigma}\Bigl(\frac{\nabla^{\sigma}\rho}{\rho}&&\Bigr)+\frac{1}{2}\frac{\nabla_{\sigma}\rho}{\rho}\frac{\nabla^{\sigma}\rho}{\rho}\Biggr)=\frac{\nabla_{\mu}\nabla^{\mu}(\rho\ln\rho)}{\rho+\rho\ln\rho} \label{VN_v1} \\ 
&&-\frac{1}{2}\frac{\nabla_{\mu}(\rho\ln\rho)\nabla^{\mu}(\rho\ln\rho)}{(\rho+\rho\ln\rho)^{2}}-\frac{\nabla_{\mu}(\rho\ln\rho)\nabla^{\mu}\rho}{(\rho+\rho\ln\rho)^{2}}. \nonumber 
\end{eqnarray}
Substituting Eq.~(\ref{VN_v0}) into Eq.~(\ref{VN_v1}) one last time, and rearranging the result, one can obtain the expression
\begin{eqnarray}
\Biggl(\nabla_{\sigma}\Bigl(&&\frac{\nabla^{\sigma}\rho}{\rho}\Bigr)+\frac{1}{2}\frac{\nabla_{\sigma}\rho}{\rho}\frac{\nabla^{\sigma}\rho}{\rho}\Biggr)=\frac{1}{\rho+\rho\ln\rho}\Biggl[\nabla_{\mu}\nabla^{\mu}(\rho\ln\rho) \nonumber \\ 
&&-\frac{1}{2}\gamma[\rho]\nabla_{\mu}(\rho\ln\rho)\nabla^{\mu}(\rho\ln\rho)\Biggr] \label{VN_v2}
\end{eqnarray}
\begin{eqnarray}
\textup{where}\;\; \gamma[\rho]=\frac{3\rho+\rho\ln\rho}{(\rho+\rho\ln\rho)^{2}}. \nonumber 
\end{eqnarray}
For relatively uniform distributions, we can presume that the entropy is significantly larger than the density $\mathfrak{S}=-\rho\ln\rho\gg\rho$, resulting in a Von Neumann entropy expression similar to that of Eq.~(\ref{Sinfo})
\begin{eqnarray}
\mathfrak{S}\nabla_{(\mu}\nabla_{\nu)}&&\mathfrak{S}-\frac{1}{2}\nabla_{(\mu}\mathfrak{S}\nabla_{\nu)}\mathfrak{S} \label{Svn} \\
&&=\frac{2}{n}g_{\mu\nu}\Bigl[\mathfrak{S}\nabla_{\sigma}\nabla^{\sigma}\mathfrak{S}-\frac{1}{2}\nabla_{\sigma}\mathfrak{S}\nabla^{\sigma}\mathfrak{S}\Bigr]. \nonumber
\end{eqnarray}
Using integration by parts, this identity can be re-expressed as
\begin{eqnarray}
\nabla_{(\mu}(\mathfrak{S}\nabla_{\nu)}&&{\mathfrak{S}})-\frac{3}{2}\nabla_{(\mu}\mathfrak{S}\nabla_{\nu)}\mathfrak{S} \nonumber \\
&&=\frac{2}{n}g_{\mu\nu}\Bigl[\nabla_{\sigma}(\mathfrak{S}\nabla^{\sigma}\mathfrak{S})-\frac{3}{2}\nabla_{\sigma}\mathfrak{S}\nabla^{\sigma}\mathfrak{S}\Bigr]. \label{Svn1}
\end{eqnarray}
We can further take advantage of the fact that $\mathfrak{S}\nabla_{\mu}\mathfrak{S}=\frac{1}{2}\nabla_{\mu}\mathfrak{S}^{2}$ to attain
\begin{eqnarray}
\nabla_{(\mu}\nabla_{\nu)}&&\mathfrak{S}^{2}-3\nabla_{(\mu}\mathfrak{S}\nabla_{\nu)}\mathfrak{S} \nonumber \\
&&=\frac{2}{n}g_{\mu\nu}\Bigl[\nabla_{\sigma}\nabla^{\sigma}\mathfrak{S}^{2}-3\nabla_{\sigma}\mathfrak{S}\nabla^{\sigma}\mathfrak{S}\Bigr]. \label{Svn2}
\end{eqnarray}
These last two steps were critical in assuring the Von Neumann entropy inherits a physical meaning. 
Generally speaking, in adopting the entropy identities of Eq.~(\ref{Sinfo})-(\ref{Svn2}), one must be careful in interpreting their corresponding uncertainty. In particular, handling entropies which are no longer discrete, rather continuous, as may be needed in field theoretic frameworks, can become quite tricky. The proper definition of entropy for continuous probability distribution functions is known as \textit{differential entropy}. The differential entropy is unphysical in every respect. It does not properly scale as discrete entropies would and contains an arbitrary infinite term when discretized from its continuous form
\begin{eqnarray}
\int_{-\infty}^{\infty}[p(x)\mathrm{ln}&&p(x)]\mathrm{d}x=\displaystyle{\lim_{\Delta\to0}}\sum_{i}\Delta p_{i}(x_{i})\mathrm{ln}\bigl(\Delta p_{i}(x_{i})\bigr) \nonumber \\
&&=\int_{-\infty}^{\infty}[p(x)\mathrm{ln}p(x)]\mathrm{d}x+\displaystyle{\lim_{\Delta\to0}}\mathrm{ln}\Delta. \label{diff_entropy_div}
\end{eqnarray}
Here, $\displaystyle{\lim_{\Delta\to0}}\mathrm{ln}\Delta\to\infty$ is a divergent term that destroys any physical meaning of the differential entropy. The only way to really overcome such a limitation in studying the entropy of continuous probabilities is by considering a reference frame for the system of interest. In that respect, many people adopt either the relative entropy or a measure known as mutual information~\cite{wehrl1978entropy}. Another possible way of defining entropy without invoking an arbitrary reference frame is by considering the fact that any partial derivative of the differential entropy effectively removes the aforementioned infinities. 
As an example, the divergence associated to the first-order partial derivative of the entropy trivially disappears 
\begin{eqnarray}
\frac{\partial{S}}{\partial{x}}&&=\displaystyle{\lim_{\Delta\to0}}\sum_{i}\Delta\Biggl[\frac{p_{i}(x_{i})\mathrm{ln}\bigl(\Delta p_{i}(x_{i})\bigr)-p_{i}(x'_{i})\mathrm{ln}\bigl(\Delta p_{i}(x'_{i})\bigr)}{|x_{i}-x'_{i}|}\Biggr] \nonumber \\ 
&&=\displaystyle{\lim_{\Delta\to0}}\sum_{i}\Delta\Biggl[\frac{p_{i}(x_{i})\mathrm{ln}\bigl(p_{i}(x_{i})\bigr)-p_{i}(x'_{i})\mathrm{ln}\bigl(p_{i}(x'_{i})\bigr)}{|x_{i}-x'_{i}|}\Biggr] \nonumber \\ 
&&=\int_{-\infty}^{\infty}\Biggl[\frac{p(x)\mathrm{ln}\bigl(p(x)\bigr)-p(x')\mathrm{ln}\bigl(p(x')\bigr)}{|x-x'|}\Biggr]. \label{cov_diff_entropy}
\end{eqnarray}
Here, $\displaystyle{\lim_{\Delta\to0}}\Biggl[\frac{\mathrm{ln}\Delta-\mathrm{ln}\Delta}{|x-x'|}\Biggr]$ was removed in transitioning from the first to the second line. All higher-order derivatives of the differential entropy similarly follow in being well-defined quantities. Therefore, the identities of Eq.~(\ref{Sinfo})-(\ref{Svn2}) are finite representations of the differential entropy so long as they are represented by their first- and second-order spatio-temporal variations. For example, the first component on the right- and left-hand sides of Eq.~(\ref{Svn}) may lead to divergences and should be studied using alternative representations, such as those of Eq.~(\ref{Svn1}) and Eq.~(\ref{Svn2}). 

In what follows, I study the information identity of Eq.~(\ref{Sinfo}), rather than its Von Neumann entropy form, to alleviate the need to deal with the squared entropy of Eq.~(\ref{Svn2}).
More specifically, I study the null- and space-like hypersurfaces of Eq.~(\ref{Sinfo}) integro-differential form. It will be shown that, for a finite system, the proposed entropy relation acts as a balancing mechanism between volumetric $n$- and $(n-1)$-forms, with $n$ being the dimensionality of the system (throughout this manuscript, we adopt $n=4$). Given the similarities of Eq.~(\ref{Sinfo}) and Eq.~(\ref{Svn}), the results which follow can easily be extended to that of the latter. The final expressions are conjectured to be holographic relations that rigorously generalize the second law of thermodynamics to its covariant form.

\section{A Second Law - Conjecture}
\subsection{A Conjecture for Null Geodesics}
In identifying the intrinsic relationship of information to spacetime, Eq.~(\ref{Sinfo}) is extended to its integro-differential form. To begin with, we study the null geodesics traversing from the enclosed spatial surface area $A$, about an $(n-1)$ dimensional null hypersurface. The null hypersurface contains normal $\hat{n}_{\mu}$ and transverse $\hat{k}^{\mu}$ vectors traversing along the null geodesics of the enclosed volumetric $n$-form. Inserting the transverse vectors $\hat{k}^{\mu}\hat{k}^{\nu}$ and applying an $n$ dimensional integral over Eq.~(\ref{Sinfo}), one arrives at the following expression
\begin{eqnarray}
\int&&\nabla_{(\mu}\nabla_{\nu)}{S}\hat{k}^{\mu}\hat{k}^{\nu}\epsilon_{abcd}-\frac{1}{2}\int\nabla_{(\mu}{S}\nabla_{\nu)}{S}\hat{k}^{\mu}\hat{k}^{\nu}\epsilon_{abcd} \nonumber \\
&&=g_{\mu\nu}\hat{k}^{\mu}\hat{k}^{\nu}\frac{2}{n}\int\Bigl[\nabla_{\sigma}\nabla^{\sigma}S-\frac{1}{2}\nabla_{\sigma}S\nabla^{\sigma}S\Bigr]\epsilon_{abcd}. \label{vacuumTI_v0}
\end{eqnarray}
Here, $\epsilon_{abcd}$ is the volumetric $n$-form. By further accounting for the fact that: (1) $g_{\mu\nu}\hat{k}^{\mu}\hat{k}^{\nu}=0$ for transverse vectors on a null hypersurface; and (2) Stoke's theorem reduces the volumetric $n$-form of $\nabla_{(\mu}\nabla_{\nu)}{S}$ to a volumetric $(n-1)$-form of the entropy flux $\nabla_{(\mu}{S}\hat{n}_{\nu)}$; one arrives at a more familiar relation of Eq.~(\ref{vacuumTI_v0}) 
\begin{eqnarray}
\int\nabla_{(\mu}{S}\hat{n}_{\nu)}\hat{k}^{\mu}\hat{k}^{\nu}\epsilon_{abc}-\frac{1}{2}\int\nabla_{(\mu}{S}\nabla_{\nu)}{S}\hat{k}^{\mu}\hat{k}^{\nu}\epsilon_{abcd}=0. \nonumber \\ \label{vacuumTI_v1}
\end{eqnarray}
Here, the volumetric $n$-form $\epsilon_{abcd}$ has been reduced to the volumetric $(n-1)$-form $\epsilon_{abc}$,
as demonstrated in Appendix B of~\cite{flanagan2000CEB}. To make the expression somewhat more intuitive, the indices of the covariant derivatives of the entropy and normal vectors can be contracted with the transverse vectors 
\begin{eqnarray}
\int\hat{S}\epsilon_{abc}=\frac{1}{2}\int\hat{S}^{2}\epsilon_{abcd}, \label{vacuumTI_v2}
\end{eqnarray}
such that $\hat{S}=\nabla_{\mu}{S}\hat{k}^{\mu}$ is the covariant entropy projected onto the tangent vector $\hat{k}^{\mu}$ and $\hat{n}_{\mu}\hat{k}^{\mu}=1$ (for null hypersurfaces). Given the covariant entropy was projected along the null geodesics, the symmetric notation around the indices are no longer necessary. For time-like and space-like geodesics the scalar contribution on the right hand side of Eq.~(\ref{vacuumTI_v0}) would remain, leaving a factor of $1/2$ which was presently removed. 

As for the second law of thermodynamics, there exists a simple argument which suggests that the entropy expressed within Eq.~(\ref{vacuumTI_v2}) is bound to increase in time. Assume there exists an $(n-1)$ dimensional spatial sphere containing a positive-definite energy density, with null geodesics traversing forward in time from its $(n-2)$ dimensional spatial boundary. Within the enclosed volumetric $n$-form (bulk of the sphere extending in time), one can presume a positive-definite covariant entropy squared $\int\hat{S}^{2}\epsilon_{abcd}\ge0$. Furthermore, for a positive-definite density, the entropy's spatial variation at the sphere's boundary is $\nabla{S}\cdot\hat{n}\propto-\nabla\rho\cdot\hat{n}\ge0$. In such a scenario, the entropy flux at the boundary of the sphere must be correspondingly positive-definite
\begin{eqnarray}
\int d^{3}x\dot{S}=\int d^{3}x[\nabla{S}\cdot\hat{n}]\;\;\to\;\;\int d^{3}x\dot{S}\ge0, \label{vacuumTI_v4}
\end{eqnarray}
resulting in the proper coherence with the second law of thermodynamics. 
Another scenario of interest is the behavior of the entropy at the $(n-2)$ dimensional spatial boundary $\Omega$ encompassing an infinitely large $(n-1)$ dimensional sphere, such that $\lim\limits_{\Omega\to\infty}\int_{\Omega}\hat{S}\epsilon_{abc}\approx0$. By recognizing that the temporal variation of the entropy squared is always positive-definite $\dot{S}^{2}\ge0$, the spatial variation of the entropy squared, integrated over spacetime, is guaranteed to be positive-definite
\begin{eqnarray}
\int d^{4}x(\nabla{S})^{2}=\int d^{4}x(\dot{S})^{2}\;\;\to\;\;\int d^{4}x(\nabla{S})^{2}\ge0. \nonumber \\ 
\label{vacuumTI_v3}
\end{eqnarray}
There is also the question of the entropy bound for which Eq.~(\ref{vacuumTI_v2}) obeys. It has already been established~\cite{buosso1999CEB,bousso2003CEB} that the covariant entropy flux along the volumetric $(n-1)$-form of the null hypersurface, starting from an enclosed $(n-2)$ spatial area $A$, must be less than or equal to $A/4\ell_{p}^{2}$, with $\ell_{p}$ being Planck's length. In assuming the dominant-energy condition, Flanagan et al.~\cite{flanagan2000CEB} established the legitimacy of a generalized CEB by using the Raychaduri's equation to prove that the area law holds for a set of geodesics of non-positive expansion starting from the surface area $A$. 

To clarify the manner by which Eq~\ref{vacuumTI_v2}, to some limit, formally satisfies the generalized CEB, one must recognize that the CKV in Eq.~(\ref{CKVlambda_v0}) articulates the symmetries of the stress-energy tensor $T_{\mu\nu}$ and that the stress energy tensor obeys Einstein's equation of state. This can be demonstrated by more carefully studying the procedure which led to the CKV. In first summing over all symmetric permutations of $\lambda$'s scalar equation of motion 
\begin{eqnarray}
g_{\mu\nu}\frac{\delta\mathcal{L}}{\delta\Omega}=\frac{1}{2}\Biggl[g\cdot\textup{Tr}\Biggl(\frac{\delta\mathcal{L}}{\delta g}\Biggr)\Biggr]_{(\mu\nu)}, \label{ckv_proof0}
\end{eqnarray}
followed by equating it to the mediating field's tensor equation of motion
\begin{eqnarray}
g_{\mu\nu}\frac{\delta\mathcal{L}}{\delta\Omega}=\frac{\delta\mathcal{L}}{\delta g^{\mu\nu}}, \label{ckv_proof1}
\end{eqnarray}
the two expressions, namely Eq.~(\ref{ckv_proof0}) and Eq.~(\ref{ckv_proof1}), can be merged to attain the sought after identity 
\begin{eqnarray}
\frac{\delta\mathcal{L}}{\delta g^{\mu\nu}}=\frac{1}{2}\Biggl[g\cdot\textup{Tr}\Biggl(\frac{\delta\mathcal{L}}{\delta g}\Biggr)\Biggr]_{(\mu\nu)}. \label{ckv_proof2}
\end{eqnarray}
Here, the indices belonging to $g_{\mu\nu}$ within the square brackets of Eq.~(\ref{ckv_proof0})-(\ref{ckv_proof2}) were removed for the purpose of simplicity. The symmetric permutation of the indices outside of the square brackets express the manner by which the exterior metric tensor is contracted with the traced expression. Given the symmetric nature of the stress-energy tensor $T_{\mu\nu}$, this procedure does nothing more than ensure the identity $T_{\mu\nu}=\frac{1}{2}T_{(\mu\nu)}$ is obeyed. In further applying the aforementioned tangent vectors, one attains 
\begin{eqnarray}
T_{\mu\nu}\hat{k}^{\mu}\hat{k}^{\nu}=\frac{1}{2}\Biggl[g\cdot\textup{Tr}\Biggl(\frac{\delta\mathcal{L}}{\delta g}\Biggr)\Biggr]_{(\mu\nu)}\hat{k}^{\mu}\hat{k}^{\nu}. \label{ckv_proof3}
\end{eqnarray}
This expression can then be shown to naturally inhibit an area law by respecting the equation of state $\delta{Q}=TdS$ which follows from Einstein's equation, where $\delta{S}$ and $\delta{Q}$ are the variations of the differential entropy and thermal energy, respectively. In considering, for example, a black hole with a local Rindler space $\mathcal{P}$ and a vanishing expansion at the horizon,
Jacobson~\cite{jacobson1995eom} showed that the heat flux in the past Rindler space $\mathcal{P}$ can be expressed as a function of the stress-energy tensor, along with an acceleration constant $\kappa$
\begin{eqnarray}
\delta{Q}=-\kappa\int_{\mathcal{H}}\tau T_{\mu\nu}\hat{k}^{\mu}\hat{k}^{\nu}\mathrm{d}\tau\mathrm{d}A. \label{Tuv_eqn}
\end{eqnarray}
Here, the domain of integration $\mathcal{H}$ is taken over a set of null generators ``inside" the past horizon of $\mathcal{P}$, with $\tau$ being the affine parameter. Jacobson's procedure is similar to that outlined in Eq.~(\ref{ckv_proof3}), exemplifying the correspondence of the CKV to the heat flux of Eq.~(\ref{Tuv_eqn}). In further ignoring the shear contribution within Raychaudhuri's equation~\cite{jacobson1995eom} 
\begin{eqnarray}
\delta{A}=-\int_{\mathcal{H}}\tau R_{\mu\nu}\hat{k}^{\mu}\hat{k}^{\nu}\mathrm{d}\tau\mathrm{d}A, \label{raych_eqn}
\end{eqnarray}
one attains a correspondence of the variation in the area on the Rindler space to the Ricci tensor. By then utilizing the Unrah temperature~\cite{unrah1976temp} within the equation of state $\delta{Q}=Td{S}=(\hbar\kappa/2\pi)\eta\delta{A}$, one arrives at a description by which $T_{\mu\nu}$ indirectly dictates the area's rate of expansion. The value of $\eta$ which allows Einstein's equation to satisfy the prefactor $8\pi G$, and as a result cohere with the CEB, is $\eta^{-1/2}=2\ell_{p}$. The CKV, projected along tangent vectors and integrated over spacetime, is then just proportional to the area times the proportionality factor $T/4\ell_{p}^{2}$. It is then apparent that any CKV attained from an action containing the Einstein-Hilbert Lagrangian, with the appropriate constant of proportionality, is bound to conform to the CEB. Furthermore, in satisfying the CEB within the proposed theory, the equation of state, and hence the first law of thermodynamics, must be obeyed, a matter which will be explored further in the next section. 

It is important to recognize that Flanagan et al.~\cite{flanagan2000CEB} proof of the CEB explicitly mentions that it holds only under the assumption of ``minimal spatio-temporal variation of the entropy." 
In this regard, a homogeneous electron gas, resulting from a relatively local scalar field, may suffice in satisfying the prior assumption of a dominant energy condition. 
Beyond this regime, a different energy condition may have to be adopted~\cite{smolin2001CEB} in discerning the validity of the entropy bound. 

The identity of Eq.~(\ref{vacuumTI_v2}) expresses a peculiar holographic semblance: it relates the entropy flux through the volumetric $(n-1)$-form, starting from an $(n-2)$ spatial area $A$, to the covariant entropy of the enclosed volumetric $n$-form. 
For null geodesics, the tangent vectors, contracted with the metric tensor within the CKV of Eq.~\ref{CKVlambda_v0}, result in the KV relation of Eq.~\ref{KVlambda}. 
Null geodesics therefore directly correspond to the field alignment scenario, whereby the conservation conditions for both the scalar and mediating are satisfied and the scalar field obeys a massless Klein-Gordon field. The following conjecture can then be made for the corresponding behavior of information 

\begin{adjustwidth}{1cm}{1cm}
	In the limit of the field alignment scenario, the covariant entropy traversing about the volumetric $(n-1)$-form of a null hypersurface, starting from an $(n-2)$ spatial area $A$, must be balanced by half times the covariant entropy squared tangent to the null geodesics of the enclosed volumetric $n$-form.
\end{adjustwidth}

The volumetric entropy squared can be perceived as equating to two times the surface entropy flux so as to account for both past and future directed light-sheets. 
For time-like geodesics, we shall see that the conformal structure of the identity becomes pertinent in characterizing its corresponding integro-differential equation.

\subsection{A Conjecture Beyond Null Geodesics}
Articulating the identity of Eq.~(\ref{vacuumTI_v0}) for time-like geodesics is just as important in discerning the behavior of fields beyond the field alignment scenario. In doing so, we instead adopt a space-like hypersurface, whereby $g_{\mu\nu}\hat{k}^{\mu}\hat{k}^{\nu}=+1$
\begin{eqnarray}
\int&&\nabla_{(\mu}\nabla_{\nu)}{S}\hat{k}^{\mu}\hat{k}^{\nu}\epsilon_{abcd}-\frac{1}{2}\int\nabla_{(\mu}{S}\nabla_{\nu)}{S}\hat{k}^{\mu}\hat{k}^{\nu}\epsilon_{abcd} \nonumber \\
&&=\frac{2}{n}\int\Bigl[\nabla_{\sigma}\nabla^{\sigma}S-\frac{1}{2}\nabla_{\sigma}S\nabla^{\sigma}S\Bigr]\epsilon_{abcd}. \label{identityTG_v0}
\end{eqnarray}
Here, tangent vectors $\hat{k}^{\mu}$ contract with the covariant derivative of the entropy, deeming the surface and volumetric integral expressions on the left-hand side different from those on the right. In applying Stokes theorem as before, one arrives at the following expression
\begin{eqnarray}
\int&&(\nabla_{\mu}{S}\hat{k}^{\mu})(\hat{n}_{\nu}\hat{k}^{\nu})\epsilon_{abc}-\frac{1}{2}\int(\nabla_{\mu}{S}\hat{k}^{\mu})(\nabla_{\nu}{S}\hat{k}^{\nu})\epsilon_{abcd} \nonumber \\
&&=\frac{1}{n}\Bigl[\int\nabla_{\sigma}S\hat{n}^{\sigma}\epsilon_{abc}-\frac{1}{2}\int\nabla_{\sigma}S\nabla^{\sigma}S\epsilon_{abcd}\Bigr]. \label{identityTG_v1}
\end{eqnarray}
By further recognizing that the contracted normal and tangent vectors along the volumetric $(n-1)$-form are zero $\hat{n}_{\mu}\hat{k}^{\mu}=0$, one can remove the covariant entropy on the right-hand side, resulting in an additional simplification
\begin{eqnarray}
\int\widetilde{S}\epsilon_{abc}=\frac{1}{2}\int\Bigl[\nabla_{\sigma}S\nabla^{\sigma}S-n\hat{S}^{2}\Bigr]\epsilon_{abcd}. \label{identityTG_v2}
\end{eqnarray}
Here, $\widetilde{S}=\nabla_{\sigma}S\hat{n}^{\sigma}$ is the entropy normal to the space-like hypersurface. This identity suggests that, unlike null geodesics, contributions of the covariant entropy tangent to the geodesic vectors of the space-like hypersurface negatively contribute to the total entropy flux through the prescribed volumetric $(n-1)$-form. 
Unlike null geodesics, time-like geodesics (of non-zero mass) no longer obey killing fields, rather conformal killing vector fields. Such CKVs indicate the presence of a spin-structure pertinent in characterizing time-like geodesics normal to a space-like hypersurface. 

For any given vector field on a curved manifold structure, its constituting normal and tangent components should allow one to locally reconstruct the original vector. The difference in the intensity of the covariant entropy squared $\nabla_{\mu}S\nabla^{\mu}S$ and the covariant entropy squared projected onto the tangent vectors of the space-like hypersurface $\hat{S}^{2}$ should then be regarded as the normal component of the covariant entropy squared $\nabla_{\mu}S\nabla^{\mu}S-\hat{S}^{2}=(\nabla_{\mu}S\hat{n}^{\mu})(\nabla_{\nu}S\hat{n}^{\nu})=\widetilde{S}^{2}$. One can use this relation to re-express Eq.~(\ref{identityTG_v2})
\begin{eqnarray}
\int\widetilde{S}\epsilon_{abc}=\frac{1}{2}\int\Bigl[\widetilde{S}^{2}-(n-1)\hat{S}^{2}\Bigr]\epsilon_{abcd}. \label{identityTG_v3}
\end{eqnarray}
It is the dis-proportionality between the tangent and normal covariant entropies that makes Eq.~(\ref{identityTG_v3}) interesting. 
In addition to the normal component of the covariant entropy squared $\int\widetilde{S}^{2}\epsilon_{abcd}$, $(n-1)$ times the tangent covariant entropy squared $\int\hat{S}^{2}\epsilon_{abcd}$ negates the total entropy flux escaping the volumetric $(n-1)$-form.

To clarify this, notice that the key difference between the identity of the null and space-like hypersurface is the persisting tangent covariant entropy squared within Eq.~(\ref{identityTG_v3}). In a scenario whereby the tangent contribution is nullified, the identity reduces to an expression analogous to that of a null hypersurface
\begin{eqnarray}
\int\widetilde{S}\epsilon_{abc}=\frac{1}{2}\int\widetilde{S}^{2}\epsilon_{abcd}. \label{identityTG_v4}
\end{eqnarray}
Here, the normal component of the covariant entropy squared plays the only prior role in characterizing a `balance of information.' In such a scenario, Eq.~(\ref{identityTG_v4}) takes exactly the same form as the null geodesics of Eq.~(\ref{vacuumTI_v2}), whereby tangent and normal vectors align. Beyond null geodesics, normal and tangent vectors differ and the tangent entropy flux appearing in Eq.~(\ref{identityTG_v3}) is unavoidable. 

Unlike null geodesics, the tangent covariant entropy contained within the $n$-form clearly reduces the normal entropy flux of time-like geodesics through the volumetric $(n-1)$-form. 
This hinders at a geometric spin-like dependence not encountered for null geodesics.
A conjecture follows for the corresponding information identity: 

\begin{adjustwidth}{1cm}{1cm}
	Beyond the field alignment scenario, the covariant entropy normal to a volumetric $(n-1)$-form of a space-like hypersurface, starting from an $(n-2)$ spatial area $A$, must be balanced by half the normal covariant entropy squared minus $(n-1)$ times the tangent covariant entropy squared of the enclosed volumetric $n$-form. 
\end{adjustwidth}

However insightful, these conjectures may lead to ill-defined results if Einstein's equation of state, used in conforming to the CEB (Eq.~(\ref{ckv_proof0})-(\ref{raych_eqn}) ), is not properly obeyed. To remedy this, I shall explore metriplectic systems in the section to follow so as to preserve the symplectic behavior of quantum mechanics and, in turn, conform to the first law of thermodynamics.

\section{A First Law}
Given a prospective second law emerges from the constraint of the proposed theory, it is only appropriate to question the implication of the identity of Eq.~(\ref{Sinfo}) on the strongly grounded assumption of energy conservation. In studying null geodesics traversing along null hypersurfaces, it was already emphasized that conforming to Einstein's equation of state, and therefore the first law of thermodynamics, is of paramount importance in attaining the CEB correspondence. After all, without the fundamental assumption of energy conservation, quantum mechanics is no longer quantum mechanics. 

In quantum mechanics, ensuring energy conservation amounts to utilizing the Poisson bracket $\{\cdot,\cdot\}$ to evolve an arbitrary operator $\hat{\mathcal{O}}$ in a symplectic manner
\begin{eqnarray}
\dot{\hat{\mathcal{O}}}=\{\hat{\mathcal{O}},\hat{\mathcal{H}}\}. \label{symplectic}
\end{eqnarray}
Although such an approach is sufficient for systems obeying solely the first law, incorporating a second law may become important in classes of systems that contain diffusion behavior. To remedy this, metriplectic systems~\cite{kaufman1984MP,morrison1986MP} were suggested as a way of introducing an additional dynamical degree of freedom while still conserving the symplectic behavior of the Hamiltonian. More specifically, Casimir invariants $\hat{\mathcal{R}}$ of the Poisson bracket are introduced which satisfy
\begin{eqnarray}
\{\hat{\mathcal{R}},\hat{\mathcal{H}}\}=0. \label{casimir}
\end{eqnarray}
The Casimir invariant $\hat{\mathcal{R}}$, heron referred to as the entropy functional, can then be used to invoke a second law; a degree of freedom which does not interfere with the first law, and can arbitrarily grow by means of some Leibniz bracket $(\cdot,\cdot)$ 
\begin{eqnarray}
(\hat{\mathcal{R}},\hat{\mathcal{R}})>0. \label{leibnizB}
\end{eqnarray}
Here, the Leibniz bracket conforms to the metric tensor and is symmetric
\begin{eqnarray}
(\hat{\mathcal{R}},\hat{\mathcal{R}})=g^{\mu\nu}\partial_{\mu}\hat{\mathcal{R}}\partial_{\nu}\hat{\mathcal{R}}. \label{leibniz}
\end{eqnarray}
So long as $\hat{\mathcal{R}}$ is positive-definite, the Leibniz bracket guarantees the inequality of Eq.~(\ref{leibnizB}). For a metriplectic system, the evolution of an arbitrary observable $\hat{\mathcal{O}}$ is then
\begin{eqnarray}
\dot{\hat{\mathcal{O}}}=\{\hat{\mathcal{O}},\hat{\mathcal{H}}\}+(\hat{\mathcal{O}},\hat{\mathcal{R}}). \label{MPbracket0}
\end{eqnarray}
The free energy $\hat{\mathcal{F}}$ can further be defined as the sum of the Hamiltonian and entropy functionals
\begin{eqnarray}
\hat{\mathcal{F}}=\hat{\mathcal{H}}+\hat{\mathcal{R}}. \label{energyF}
\end{eqnarray}
The metriplectic bracket $\left\langle\left\langle\cdot,\cdot\right\rangle\right\rangle$ can then be used to simplify Eq.~(\ref{MPbracket0})
\begin{eqnarray}
\dot{\hat{\mathcal{O}}}=\left\langle\left\langle\hat{\mathcal{O}},\hat{\mathcal{F}}\right\rangle\right\rangle. \label{MPbracket1}
\end{eqnarray}
The free energy is the portion of energy available to do thermodynamic work and is therefore of important physical significance. Even in the use of effective potentials within renormalization theory, free energy functionals have become extremely useful in attaining the expectation value of quantum fields~\cite{peskin2018qft}, whereby: renormalization terms are incorporated within an effective action, quantum fluctuations are replaced by thermal fluctuations, and the resulting free energy is minimized. In the framework of metriplectic systems, the free energy functional and its corresponding bracket assure that the symplectic behavior of a Hamiltonian is not altered by the corresponding entropy functional. The identities which follow from the metriplectic bracket are
\begin{eqnarray}
&&\left\langle\left\langle\hat{\mathcal{H}},\hat{\mathcal{F}}\right\rangle\right\rangle=0 \;\;\to\;\; \{\hat{\mathcal{H}},\hat{\mathcal{R}}\}=0 \nonumber \\
&&\left\langle\left\langle\hat{\mathcal{R}},\hat{\mathcal{F}}\right\rangle\right\rangle>0 \;\;\to\;\; (\hat{\mathcal{R}},\hat{\mathcal{R}})>0. \label{MPbracket2} 
\end{eqnarray}
These identities imply that the entropy functional effectively diffuses the information along the Casimir invariants of the Hamiltonian, resulting in an overall increase in entropy while still allowing the Hamiltonian to conserve energy. In what follows, the Hamiltonian and entropy functionals of the proposed theory will be more concretely defined. I will thereafter rigorously derive the Poisson bracket of Eq.~(\ref{casimir}) and exemplify that, in the field alignment scenario, one simply retrieves Einstein's energy-mass relation.

\subsection{Beyond Hamiltonians}
The Hamiltonian and entropy functionals described in the previous section can be rigorously derived in the scalar field representation. Although such contributions can be defined by simply taking the energy component of the stress-energy tensor $T_{00}$, a procedure for articulated the free energy functional will be exemplified in what follows. Typically, in characterizing the Hamiltonian of a purely symplectic system composed of a complex scalar field $\varphi$, the conjugate momentum $\pi=\frac{\partial\mathcal{L}_{s}}{\partial{\dot{\varphi}}}$ can be used to arrive at the following expression
\begin{eqnarray}
\hat{\mathcal{H}}=(\pi\dot{\varphi}+\pi^{*}\dot{\varphi}^{*})-\mathcal{L}. \label{freeEv0}
\end{eqnarray}
Such a Hamiltonian suffices in considering first-order temporal variations of the scalar field. In instead accounting for $N_{d}$ higher-order temporal derivatives of the scalar field, the free energy functional takes the form
\begin{eqnarray}
\hat{\mathcal{F}}=\sum_{\ell}^{N_{d}}(\pi_{\ell}\partial_{t}^{\ell}\varphi+\pi_{\ell}^{*}\partial_{t}^{\ell}\varphi^{*})-\mathcal{L}. \label{freeEv1}
\end{eqnarray}
Here, $\partial_{t}^{\ell}$ refers to the $\ell$th order temporal derivative and the $\ell$th order conjugate momentum is simply $\pi_{\ell}=\frac{\partial\mathcal{L}_{\lambda}}{\partial{(\partial_{t}^{\ell}\varphi)}}$. Given the Lagrangian of Eq.~(\ref{vacuaA}), we limit ourselves to second-order temporal derivatives. To simplify matters, we handle the Lagrangian for the scalar field $\mathcal{L}_{s}$ and mediating field $\mathcal{L}_{\lambda}$ separately, such that the resulting energy functionals are $\hat{\mathcal{H}}$ and $\hat{\mathcal{R}}$, respectively. Starting with the Hamiltonian $\hat{\mathcal{H}}$, we see that it is no different from the usual expression, up to a conformal factor
\begin{eqnarray}
\hat{\mathcal{H}}&=&(\pi_{1}\dot{\varphi}+\pi_{1}^{*}\dot{\varphi}^{*})-\mathcal{L}_{s}-\mathcal{L}_{\lambda} \nonumber \\
&=&\frac{1}{2}\Omega^{2}\Bigl[(\dot{\varphi})^{2}+(\nabla{\varphi})^{2}+\frac{m^{2}}{\hbar^{2}}(\varphi)^{2}\Bigr]. \label{freeEv2}
\end{eqnarray}
Here, $\mathcal{L}_{\lambda}=0$ due to the expression attained in varying the Lagrangian with respect to the mediating field 
\begin{eqnarray}
\frac{\delta\mathcal{L}_{\lambda}}{\delta\lambda}=0\;\;\to\;\;\Bigl[\ln\Omega^{2}-\frac{\hbar^{2}}{m^{2}}\frac{\Box\sqrt{\rho}}{\sqrt{\rho}}\Bigr]=0. \label{freeEv3}
\end{eqnarray}
The entropy expression can then be determined by expanding the quantum potential into its scalar field form within the constraint of Eq.~(\ref{vacuaA}). 
\begin{eqnarray}
\frac{\Box\sqrt{\rho}}{\sqrt{\rho}}=\frac{1}{2}&&\Bigl[\frac{\Box\varphi^{*}}{\varphi^{*}}+\frac{\nabla_{\mu}\varphi^{*}}{\varphi^{*}}\frac{\nabla^{\mu}\varphi}{\varphi}+\frac{\Box\varphi}{\varphi}\Bigr] \nonumber \\
&&-\frac{1}{4}\Bigl[\frac{\nabla_{\mu}\varphi^{*}}{\varphi^{*}}\frac{\nabla^{\mu}\varphi^{*}}{\varphi^{*}}+\frac{\nabla_{\mu}\varphi}{\varphi}\frac{\nabla^{\mu}\varphi}{\varphi}\Bigr]. \label{freeEv4}
\end{eqnarray}
The first-order conjugate momentums associated to $\mathcal{L}_{\lambda}$ are given by
\begin{eqnarray}
\pi_{1}\dot{\varphi}=\frac{\partial\mathcal{L}_{\lambda}}{\partial\dot{\varphi}}\dot{\varphi}&=&\frac{1}{2}\Bigl[\frac{\dot{\varphi}^{*}}{\varphi^{*}}\frac{\dot{\varphi}}{\varphi}-\frac{\dot{\varphi}}{\varphi}\frac{\dot{\varphi}}{\varphi}\Bigr]\lambda \nonumber \\
\pi_{1}^{*}\dot{\varphi}^{*}=\frac{\partial\mathcal{L}_{\lambda}}{\partial\dot{\varphi}^{*}}\dot{\varphi}^{*}&=&\frac{1}{2}\Bigl[\frac{\dot{\varphi}^{*}}{\varphi^{*}}\frac{\dot{\varphi}}{\varphi}-\frac{\dot{\varphi}^{*}}{\varphi^{*}}\frac{\dot{\varphi}^{*}}{\varphi^{*}}\Bigr]\lambda. \label{freeEv5}
\end{eqnarray}
The second-order conjugate momentums then take the form
\begin{eqnarray}
\pi_{2}\ddot{\varphi}=\frac{\partial\mathcal{L}_{\lambda}}{\partial\ddot{\varphi}}\ddot{\varphi}&=&\frac{1}{2}\frac{\ddot{\varphi}}{\varphi}\lambda \nonumber \\
\pi_{2}^{*}\ddot{\varphi}^{*}=\frac{\partial\mathcal{L}_{\lambda}}{\partial\ddot{\varphi}^{*}}\ddot{\varphi}^{*}&=&\frac{1}{2}\frac{\ddot{\varphi}^{*}}{\varphi^{*}}\lambda. \label{freeEv6}
\end{eqnarray}
In pairing the first- and second-order conjugate momentums together, one gets the following expressions
\begin{eqnarray}
(\pi_{1}\dot{\varphi}+\pi_{1}^{*}\dot{\varphi}^{*})&=&\frac{\dot{\varphi}^{*}}{\varphi^{*}}\frac{\dot{\varphi}}{\varphi}\lambda-\frac{1}{2}\frac{\dot{\rho}}{\rho}\dot{\lambda} \nonumber \\
(\pi_{2}\ddot{\varphi}+\pi_{2}^{*}\ddot{\varphi}^{*})&=&\frac{1}{2}\frac{\ddot{\rho}}{\rho}\lambda-\frac{\dot{\varphi}^{*}}{\varphi^{*}}\frac{\dot{\varphi}}{\varphi}\lambda. \label{freeEv7}
\end{eqnarray}
In merging these expression into the entropy functional, one arrives at a final expression
\begin{eqnarray}
\hat{\mathcal{R}}&=&(\pi_{1}\dot{\varphi}+\pi_{1}^{*}\dot{\varphi}^{*})+(\pi_{2}\ddot{\varphi}+\pi_{2}^{*}\ddot{\varphi}^{*}) \nonumber \\
&=&-\frac{\hbar^{2}}{2m^{2}}\frac{1}{\rho}\Bigl[\dot{\rho}\dot{\lambda}-\ddot{\rho}\lambda\Bigr]. \label{freeEv8}
\end{eqnarray}
Here, the Lagrangian was not inserted into the expression again, given it was already accounted for within the Hamiltonian of Eq.~(\ref{freeEv2}). Finally, placing both the Hamiltonian and entropy functions into $\hat{\mathcal{F}}$ results in a complete expression of the free energy
\begin{eqnarray}
\hat{\mathcal{F}}&=&\hat{\mathcal{H}}+\hat{\mathcal{S}} \label{freeEv9} \\
&=&\frac{1}{2}\Omega^{2}\Bigl[(\dot{\varphi})^{2}+(\nabla{\varphi})^{2}+\frac{m^{2}}{\hbar^{2}}(\varphi)^{2}\Bigr]-\frac{\hbar^{2}}{2m^{2}}\frac{1}{\rho}\Bigl[\dot{\rho}\dot{\lambda}-\ddot{\rho}\lambda\Bigr]. \nonumber
\end{eqnarray}
This expression can then be used within the metriplectic brackets of Eq.~(\ref{MPbracket2}) to assure that the coherent behavior of $\lambda$ and $\varphi$ conserve energy and obey a second law. Assuring that $\{\hat{\mathcal{H}},\hat{\mathcal{H}}\}=0$ and $(\hat{\mathcal{R}},\hat{\mathcal{R}})>0$ is quite trivial. The former is know from the standard quantum field theory while the latter holds as long as the mediating field $\lambda$ is real-valued. A real-valued mediating field comes as no surprise given its governing equation of motion (Eq.~(\ref{Slambda}) ) is density-dependent. It should also be emphasized that, given $\lambda$ is a Lagrange multiplier, characterizing its free energy functional and associated brackets is meaningless. In the subsection to follow, I will demonstrate that the less trivial Poisson bracket $\{\hat{\mathcal{H}},\hat{\mathcal{R}}\}=0$ simply results in Einstein's energy-mass relation in the limit of the field alignment scenario.

\subsection{Poisson Bracket}
The Poisson bracket for complex scalar fields takes the form
\begin{eqnarray}
\{\hat{\mathcal{H}},\hat{\mathcal{R}}\}=\int{d}^{3}x\Bigl[\frac{\partial\hat{\mathcal{H}}}{\partial\varphi^{*}}\frac{\partial\hat{\mathcal{R}}}{\partial\pi}-\frac{\partial\hat{\mathcal{R}}}{\partial\varphi}\frac{\partial\hat{\mathcal{H}}}{\partial\pi^{*}}\Bigr]=0. \label{bracketP0}
\end{eqnarray}
Of these different variations, the simplest ones are those that apply to the Hamiltonian, which take the form
\begin{eqnarray}
\frac{\partial\hat{\mathcal{H}}}{\partial\varphi^{*}}&=&\frac{1}{2}\Omega^{2}\frac{m^{2}}{\hbar^{2}}\varphi \nonumber \\
\frac{\partial\hat{\mathcal{H}}}{\partial\pi^{*}}&=&\frac{1}{2}\Omega^{2}\dot{\varphi}. \label{bracketP1}
\end{eqnarray}
Attaining the variation of the entropy functional is less trivial. To begin with, the entropy functional $\hat{\mathcal{R}}$ can be re-expressed in terms of the scalar field
\begin{eqnarray}
\frac{1}{\varphi^{*}\varphi}&&\Bigl[\frac{\partial}{\partial{t}}(\varphi^{*}\varphi)\dot{\lambda}-\frac{\partial^{2}}{\partial{t}^{2}}(\varphi^{*}\varphi)\lambda\Bigr]= \nonumber \\
&&=\Bigl[\frac{\dot{\varphi}^{*}}{\varphi^{*}}+\frac{\dot{\varphi}}{\varphi}\Bigr]\dot{\lambda}-\Bigl[\frac{\ddot{\varphi}^{*}}{\varphi^{*}}+2\frac{\dot{\varphi}^{*}}{\varphi^{*}}\frac{\dot{\varphi}}{\varphi}+\frac{\ddot{\varphi}}{\varphi}\Bigr]\lambda. \label{bracketP2}
\end{eqnarray}
In then varying this expression with respect to the scalar field and conjugate momentum, one arrives at
\begin{eqnarray}
\frac{\partial\hat{\mathcal{R}}}{\partial\varphi}&=&-\frac{1}{2}\frac{\hbar^{2}}{m^{2}}\frac{1}{\rho}\Bigl[\frac{\dot{\varphi}}{\varphi}\varphi^{*}\dot{\lambda}-\frac{\ddot{\varphi}}{\varphi}\varphi^{*}\lambda-2\frac{\dot{\varphi}}{\varphi}\dot{\varphi}^{*}\lambda\Bigr] \nonumber \\
\frac{\partial\hat{\mathcal{R}}}{\partial\pi}&=&\frac{1}{2}\frac{\hbar^{2}}{m^{2}}\frac{1}{\rho}\Bigl[\varphi^{*}\dot{\lambda}-2\dot{\varphi}^{*}\lambda\Bigr]. \label{bracketP3}
\end{eqnarray}
After inserting Eq.~(\ref{bracketP1})-(\ref{bracketP3}) into the Poisson bracket $\{\hat{\mathcal{H}},\hat{\mathcal{R}}\}$, the following expression is attained
\begin{eqnarray}
\int{d}^{3}x&&\Bigl[\Omega^{2}\mathcal{Q}(\varphi,\dot{\varphi};\ell_{c})\dot{\lambda}\Bigr] \label{bracketP4} \\
&&=\int{d}^{3}x\Bigl[\Omega^{2}\Bigl(\frac{\dot{\varphi}}{\varphi}\frac{\ddot{\varphi}}{\varphi}+2\frac{\dot{\varphi}^{*}}{\varphi^{*}}\mathcal{Q}(\varphi,\dot{\varphi};\ell_{c})\Bigr)\lambda\Bigr]. \nonumber
\end{eqnarray}
Where the Compton wavelength $\ell_{c}=\hbar/m$ and the function $\mathcal{Q}$ takes the form
\begin{eqnarray}
\mathcal{Q}(\varphi,\dot{\varphi};\ell_{c})=\Bigr(\ell_{c}^{-2}+\frac{\dot{\varphi}}{\varphi}\frac{\dot{\varphi}}{\varphi}\Bigl). \label{bracketP5} 
\end{eqnarray}
At this point, this expression has no physical intuition, but one can assume that, in the limit of the field alignment scenario, the Poisson bracket should be more intuitive. For this reason we express the variation in entropy functional in the $\lambda\to\rho$ limit, so as to arrive at
\begin{eqnarray}
\frac{\partial\hat{\mathcal{R}}}{\partial\varphi}&=&\frac{1}{2}\frac{\hbar^{2}}{m^{2}}\partial_{t}\Bigl(\frac{\varphi^{*}\dot{\varphi}}{\varphi}\Bigr)=\frac{1}{2}\frac{\hbar^{2}}{m^{2}}\partial_{t}\Bigl(\varphi^{*}\partial_{t}\ln{\varphi}\Bigr) \nonumber \\
\frac{\partial\hat{\mathcal{R}}}{\partial\pi}&=&\frac{1}{2}\frac{\hbar^{2}}{m^{2}}\frac{1}{\varphi}(\varphi^{*}\dot{\varphi}-\dot{\varphi}^{*}\varphi)=\frac{\hbar^{2}}{m^{2}}\frac{i}{\hbar}\frac{1}{\varphi}\mathcal{E}. \label{bracketP6}
\end{eqnarray}
The second of these two expression simply results in the energy density $\mathcal{E}=\rho{E}$ once multiplied by $\partial\mathcal{H}/\partial\varphi^{*}$ within the Poisson bracket. It is relieving that such a complicated expression simplifies to a familiar quantity. Varying the entropy functional with respect the scalar field, on the other hand, results in a highly nontrivial expression. We can remedy this difficulty by resorting to the Bohmian picture $\varphi=\sqrt{\rho}e^{iS/\hbar}$, whereby the density $\sqrt{\rho}$ and phase $S$ of the scalar field are partitioned
\begin{eqnarray}
\partial_{t}\Bigl(\varphi^{*}&&\partial_{t}\ln{\varphi}\Bigr)=\partial_{t}\Bigl[(\partial_{t}\sqrt{\rho}+\frac{i}{\hbar}\partial_{t}S)e^{-iS/\hbar}\Bigr] \nonumber \\
&&=\Bigl[\frac{\partial_{t}^{2}\sqrt{\rho}}{\sqrt{\rho}}+\frac{i}{\hbar}\partial_{t}^{2}S+\frac{1}{\hbar^{2}}\partial_{t}S\partial_{t}S\Bigr]\varphi^{*}. \label{bracketP7}
\end{eqnarray}
We can further utilize the fact that, in the limit of the field alignment scenario, the scalar field conforms to a massless Klein-Gordon field, resulting in solutions which are temporally dependent only up to a phase. Any time derivatives of the energy are then bound to be null $\partial_{t}(\partial_{t}S)=\partial_{t}E=0$. In further substituting the gauge connection derived in~\cite{gabay2020eom} $\widetilde{D}_{t}^{2}\varphi^{*}/\varphi^{*}=(\partial_{t}+\frac{i}{\hbar}\partial_{t}S)^{2}\varphi^{*}/\varphi^{*}=\partial_{t}^{2}\sqrt{\rho}/\sqrt{\rho}$, one has
\begin{eqnarray}
\partial_{t}\Bigl(\varphi^{*}\partial_{t}\ln{\varphi}\Bigr)&&=\Bigl[\widetilde{D}_{t}^{2}+\frac{i}{\hbar}\partial_{t}^{2}S+\frac{1}{\hbar^{2}}\partial_{t}S\partial_{t}S\Bigr]\varphi^{*} \nonumber \\
&&=\frac{1}{\hbar^{2}}E^{2}\varphi^{*}. \label{bracketP8}
\end{eqnarray}
Here, $\widetilde{D}_{t}^{2}\varphi^{*}\approx0$ in the limit of the field alignment scenario. 
Once inserted into the Poisson bracket $\{\hat{\mathcal{H}},\hat{\mathcal{R}}\}$, the expression begins to appear more sensible
\begin{eqnarray}
\frac{1}{2\hbar^{2}}E^{2}\int{d}^{3}x\Bigl[{\varphi}^{*}\dot{\varphi}\Bigr]=\frac{m^{2}}{\hbar^{2}}\frac{i}{\hbar}\int{d}^{3}x\mathcal{E}. \label{bracketP9}
\end{eqnarray}
Equation~(\ref{bracketP9}) can further be simplified by recognizing that, in the limit of the field alignment scenario, $\varphi^{*}\dot{\varphi}=\frac{i}{\hbar}\rho\partial_{t}S+\frac{1}{2}\partial_{t}\rho\approx\frac{i}{\hbar}\rho\partial_{t}S=\frac{i}{\hbar}\rho E=\frac{i}{\hbar}\mathcal{E}$, resulting in a familiar relation
\begin{eqnarray}
\frac{1}{2\hbar^{2}}E^{2}\int{d}^{3}x\Bigl[{\varphi}^{*}\dot{\varphi}\Bigr]&=&\frac{m^{2}}{\hbar^{2}}\frac{i}{\hbar}\int{d}^{3}x\mathcal{E} \nonumber \\
\frac{1}{2}{E}^{2}\frac{i}{\hbar}\int{d}^{3}x\mathcal{E}&=&m^{2}\frac{i}{\hbar}\int{d}^{3}x\mathcal{E} \nonumber \\
{E}^{2}&=&2m^{2}. \label{bracketP10}
\end{eqnarray}
In the last line, we have nothing more than Einstein's energy-mass relation up to a factor of $\sqrt{2}$. This pre-factor for the mass has appeared several times in others sectors of the proposed theory~\cite{our_cosmo_paper,gabay2020eom} and the reason for it is not yet clear. This profound result suggests that once the local invariance is retrieved, the entropy functional enforces the Hamiltonian to conform to the energy-mass relation, a mechanism which is not present in the current theory of quantum mechanics.

\section{Discussion}
The proposed entropy relation suggests that the conformal nature of the killing field only vanishes in the limit of the field alignment scenario. Beyond this regime, the mass plays a prior role and its associate time-like geodesics inherit a `geometric spin'-like nature not present in scalar field theories. 

This peculiar correspondence can best be exemplified by studying the relationship between a CKV and Conformal Killing Spinor (CKS)~\cite{rajaniemi2006CKS}. CKSs articulate the correspondence between spinor $\Psi$ and tangent vector $\mathcal{J}$ fields. More specifically, the Lie derivative $\nabla_{\mathcal{J}}$ of a spinor $\Psi$ along the vector field $\mathcal{J}$ is equal to the Dirac operator applied onto the spinor field and then projected onto $\mathcal{J}$. This relationship can best be defined by the Penrose operator~\cite{penrose1967poper,penrose1984poper}
\begin{eqnarray}
\mathcal{P}\Psi=(\nabla_{\mathcal{J}}+\frac{1}{n}\mathcal{J}\cdot\slashed{\nabla})\Psi=0. \label{CKSeq}
\end{eqnarray}
Here, the Dirac operator $\slashed{\nabla}=\gamma_{\mu}\nabla^{\mu}$ adopts the Dirac slash notation, with $\gamma_{\mu}$ being the Dirac matrices. Rajaniemi \textit{et al}.~\cite{rademacher1991CKS,rajaniemi2006CKS} rigorously defined the relationship between CKVs and CKSs by suggesting that for every set of spinors $\Psi$,$\mathcal{X}$ there exists a conformal Killing vector field $\mathcal{J}_{\mu}$. In expressing the divergence of the vector field $\mathcal{J}_{\mu}$ as the spin-invariant inner product $(\cdot,\cdot)$ of $\Psi$ and $\mathcal{X}$
\begin{eqnarray}
\nabla_{\mu}\mathcal{J}^{\mu}\approx\lim\limits_{\mathcal{X}\to\Psi}\Bigl((\Psi,\slashed{\nabla}\mathcal{X})+(\slashed{\nabla}\Psi,\mathcal{X})\Bigr), \label{spin_innerproduct}
\end{eqnarray}
a correspondence can be found between the divergence of the vector field $\mathcal{J}^{\mu}$ and the Dirac current in the special case of $\mathcal{X}=\Psi$. This then results in the following CKV for $\mathcal{J}^{\mu}$
\begin{eqnarray}
\nabla_{(\mu}\mathcal{J}_{\nu)}=\frac{2}{n} g_{\mu\nu}\nabla_{\sigma}\mathcal{J}^{\sigma}. \label{CKVdirac}
\end{eqnarray}
It then becomes apparent that their exists an intrinsic relationship between the CKS of Eq.~(\ref{CKSeq}) and the CKV of Eq.~(\ref{CKVdirac}). The spacetime symmetry of the Dirac spinor resembles a CKV taking the form of the Dirac current. 

Although such a spin correspondence is incredibly luring, one must recognize that there exists a key difference between the identity of Eq.~(\ref{CKVlambda_v0}) and Eq.~(\ref{CKVdirac}): a CKS results in a CKV that is intrinsically dependent on the phase of the field variable, not the density. There is no correct way of retrieving a phase-dependent CKV from one which is intrinsically dependent on the density. 

In studying the CKV of Eq.~(\ref{CKVlambda_v0}), one could then acquire that there exists a `geometric spin' articulated by a corresponding CKS. The spin referred to here is not one associated to a field variable's change in phase and is therefore not the commonly adopted spin, rather an entity attributed to the spatio-temporal variations of the energy density. Therefore, for any nonconforming configuration of the mediating field ($\lambda\ne\rho$), one could conceive a fermionic-like particle with an underlying spin-structure. Furthermore, in the limit of a field alignment scenario ($\lambda\to\rho$), a massless particle traversing along the null geodesics of the metric could be interpreted as a bosonic-like particle. This therefore implies the characterization of either a fermionic- or bosonic-like particle in a theory whereby a mediating field $\lambda$ dictates the existence or non-existence, respectively, of the particle's corresponding geometric spin. 

In light of the early works of Dirac~\cite{dirac1937constants}, the notion of an equation of motion (Eq.~(\ref{Slambda}) ) that dictates a particle's spin by the quantity of its mass is conceptually luring. Here, the configuration of the density seems to largely dictate the existence of a geometric spin structure. 
The identity of Eq.~(\ref{KVlambda}) suggests that small spatio-temporal variations of the particle density enforce the particle to obey isometries, rather than diffeomorphisms, of the metric. This implies that a boson of spin zero can only be characterized in the limit of the field alignment scenario, whereby the mass vanishes from the scalar field's governing equation of motion. Given that all bosons we encounter in nature are massless propagators, this comes as no surprise. The photon is an excellent example: it contains a constant speed in time and a spatial helicity. Such behavior is plausible for a density that is uniform in time and local in space. The local character in space results in the helicity of the photon while the null temporal variations $\nabla_{t}\sqrt{\rho}=0$ in a constant speed of light. The covariant entropy relation of eq.~\ref{identityTG_v3} therefore provides an interesting relationship between the non(quasi)-local character of the particle density and its inherent geometric spin.

\section{Summary \& Outlook}
In the course of this manuscript, a metriplectic structure was proposed as a way of extending the symplectic theory of quantum mechanics to one which additionally acts on its Casimir invariants. The Casimir invariants were suggested to correspond to a second law of thermodynamics, whereby the behavior of information pertaining to matter was described by a relation resembling that of a density-dependent CKV. The integro-differential form of this covariant entropy relation was then shown to result in a holographic correspondence between the covariant entropy of volumetric $(n-1)$- and $n$-forms for null and time-like geodesics, starting from an $(n-2)$-spatial area $A$. The resulting entropy flux was conjectured to align with CEB so long as a Einstein's equation of state and, by extension, the first law of thermodynamics was obeyed. This was ensured by demonstrating that the Poisson bracket of the Hamiltonian commuted with the derived entropy functional. Furthermore, in the field alignment scenario, the Poisson bracket was shown to elegantly result in Einstein's energy-mass relation. 

It is the author's belief that a more rigorous proof must be provided to insure that the proposed entropy relation truly conforms to the CEB for null geodesics. One possible route would be to attempt to confirm the correspondence for a black hole with a local Rindler space. Exemplifying the entropy relation's validity in the quantum limit is just as important and doing so will be left for future works.

\appendix
\section{Killing Fields} \label{sec:app1}
KV fields articulate isometries, indicating particular spacetime symmetries which are invariant to the metric $g_{\mu\nu}$. The KV for a killing vector $K_{\mu}$ can be expressed as
\begin{eqnarray}
\nabla_{(\mu}K_{\nu)}=0. \label{KVeq}
\end{eqnarray}
Conceptually, the invariance of the KV to the metric results in worldlines for a particle to traverse. CKVs are different from KVs in that they indicate vector fields associated to general diffeomorphisms~\cite{wald1987book,weinberg1972book}. Unlike KVs, as one traverses along a CKV, the metric $g_{\mu\nu}$ is no longer left invariant, rather the change implicated by the coordinate transformation is proportional to the divergence of the killing vector field $K_{\mu}$
\begin{eqnarray}
\nabla_{(\mu}K_{\nu)}=\frac{2}{n} g_{\mu\nu}\nabla_{\sigma}K^{\sigma}. \label{CKVeq}
\end{eqnarray}
CKVs have an interesting way of collapsing back to isometries once null geodesics $u^b$ are considered. This can be seen by contracting Eq.~(\ref{CKVeq}) with $u^b$
\begin{eqnarray}
u^{a}\nabla_{a}(K_{b}u^{b})=\frac{1}{n} \nabla_{a}K^{a}u_{b}u^{b}=0. \label{CKVeq_null}
\end{eqnarray}
For null geodesics $u_{b}u^{b}=0$ and the KV equation (Eq.~(\ref{KVeq}) ) appears to be a subset of the CKV equation (Eq.~(\ref{CKVeq}) ). For timelike or spacelike geodesics, the conformal killing field can no longer be deemed null, rather must preserve the conformal structure of the metric. 

CKVs can also be intuitively characterized by a higher rank identity known as the killing tensor (KT) field~\cite{rani2003KT}. KTs are higher rank extension of KVs, and therefore also impose isometries inherit to the metric for tensor ranks greater than one
\begin{eqnarray}
\nabla_{(a}\mathcal{T}_{bc)}=0. \label{KTeq}
\end{eqnarray}
They physically prescribe worldvolumes, rather than worldlines, which can analogously be used to describe the flow of a nonlocally displaced object. The relationship between a KT field and a CKV can best be defined as
\begin{eqnarray}
\mathcal{T}_{ab}=K_{a}K_{b}-\frac{2}{n}g_{ab}K^{c}K_{c}. \label{CKVtoKT}
\end{eqnarray}
For vectors $u^{a}$ and $u^{b}$ tangent to affinely parameterized geodesics, the isometry of the Killing tensor assures
\begin{eqnarray}
u^{m}\nabla_{m}(\mathcal{T}_{ab}u^{a}u^{b})=0. \label{KTflow}
\end{eqnarray}
Perceiving CKVs as a special case of KTs is intuitively significant due to their conceptual appeal in, like KVs, giving rise to constants of motion. Although KTs will not be used in what follows, it is worth noting that, henceforth, any analysis done on Eq.~(\ref{CKVlambda_v0}) can be extended to the killing tensor identity of Eq.~(\ref{KTeq}) using Eq.~(\ref{CKVtoKT}).

\end{document}